# A new experimental set-up for aerosol stability investigations in microgravity conditions

Charles Graziani[1], Mathieu Nespoulous[2], Renaud Denoyel[3], Stephan Fauve[4], Christian Chauveau[5], Luc Deike[6], Mickaël Antoni[7]

[1]charles.graziani@univ-amu.fr, Aix-Marseille Univ, CNRS, MADIREL, Marseille, France
[2]mathieu.nespoulous@univ-amu.fr, Aix-Marseille Univ, CNRS, MADIREL, Marseille, France
[3]renaud.denoyel@univ-amu.fr, Aix-Marseille Univ, CNRS, MADIREL, Marseille, France
[4]fauve@lps.ens.fr, LP ENS-Paris, France,
[5]christian.chauveau@cnrs-orleans.fr, CNRS–ICARE, University Orléans, France
[6]ldeike@princeton.edu, MAE/HMEI, Princeton University, USA
[7]m.antoni@univ-amu.fr, Aix-Marseille Univ, CNRS, MADIREL, Marseille, France

**Abstract**
The temporal and spatial evolution of dispersed media is a fundamental problem in a wide range of physicochemical systems, such as emulsions, suspensions and aerosols. These systems are multiphasic and involve compounds of different densities. They are therefore subject to the influence of gravity which determines the sedimentation rate of their dispersed phase. This effect can be dominant and prevent a detailed study of the phenomena occurring between the constituents themselves, such as the coalescence of drops in emulsions, the evaporation of droplets or the flocculation in suspensions. In this context, the Centre National d'Etudes Spatiales (CNES) has recently supported the development of a new instrument to produce populations of droplets, a few micrometers in radius, under controlled conditions with the objective of allowing a detailed study of their properties in microgravity conditions. The principle of this instrument is to generate, by a fast compression/expansion of air, populations of water droplets and to track their evolution by optical scanning tomography in transmission mode within a volume of approximately 2 mm$^3$. Parabolic flight experiments have shown the possibility to generate and accurately follow the evolution of populations of several hundred droplets for more than 20 seconds. The first experimental results show that it is possible to study their evaporation kinetics or their motion when imposing Von Karman swirling flows. This work is part of the AEROSOL project of DECLIC-EVO supported by CNES and aims to help the understanding of cloud microphysics which remains a critical open problem in the context of global warming.

**Introduction**

An aerosol is a multiphase system in which at least one of the phases is in a dispersed form. These phases can be liquid or solid and their simultaneous evolution within the surrounding gas involves mechanisms at the molecular and macroscopic scales. Like emulsions or nanofluids, aerosols are dispersions with large exchange surfaces which give them specific physico-chemical properties, making them central to fundamental and applied fields. Aerosols can be of natural or anthropogenic origin. Fog, dust, clays, forest exudates are natural examples while smog, fine particles produced by internal combustion engines or building heating devices are all anthropogenic.

Aerosols are complex media, sensitive to temperature and pressure conditions, sometimes chemically active. They are naturally subject to the influence of gravity, which determines the settling speed of the particles they contain[1][2]. Aerosols are used in the pharmaceutical[3] and cosmetic industries (inhalers, sprays)[4] in metallurgy (spray drying)[5] and in agriculture (pesticides)[6]. From an academic point of view, they are still the subject of intense research activity. The description of their evolution is for example a major issue in climatology, especially for the understanding of cloud microphysics[7][8][9]. This problem motivated CNES to support the development of a new experimental facility for conducting experiments in parabolic flight conditions (Figure 1).

A relevant study of cloud microphysics requires well controlled conditions, especially in temperature and pressure. In clouds, droplets are made of water and constitute metastable systems produced by a rapid condensation (actually liquefaction) in a gaseous medium in which the relative humidity is sufficiently high. Despite their small size, cloud droplets are subject to the acceleration of the earth's gravity field. Their limiting velocity in the atmosphere was measured more than 70 years ago with values typically between 1 cm/s and 100 cm/s for radii ranging between 1 and 100 μm[10]. Other measurements carried out under well controlled experimental conditions (20°C, 1 atm, 50% relative humidity) show that droplets of 80 µm diameter exhibit a drainage velocity that reaches 150 cm/s[11]. Empirical laws have been proposed and give values in good coherence with these observations and with Stokes law when the diameter of the drops becomes lower than 50 µm[2]. Measurements have confirmed these settling velocities and demonstrated the key contribution of gravity in the evolution of aerosols.

In the context of cloud microphysics, this raises the question of scale separation. The characteristic times associated with the interaction between water droplets, such as coalescence, can be very short when compared to those associated with gravity. When droplets radii reach few micrometers, gravity becomes dominant and complicates studies of their mutual interactions. Performing experiments in microgravity is therefore essential for a detailed understanding of the mechanisms governing the evolution of aerosols. The reduced gravity of parabolic flights has the great advantage of decoupling buoyancy from capillary phenomena and, in the case of aerosols, of making possible an original description of the interaction mechanisms involved in a droplet population. In such conditions, it becomes indeed possible to measure the damping time of the flows created during the aerosol formation or the evolution of the droplet size distribution.

The aerosols investigated hereafter consist of populations of water droplets with radius of a few micrometres evolving in air. They are thus assimilable to fogs. From an experimental point of view, the air used is not inseminated with solid particles[12]. The one used in the experiments is not purified and therefore naturally contains particles favouring heterogeneous nucleation. This is the reason why the term 'aerosol' will be used hereafter to designate the populations of droplets under consideration. Investigations on such a 'canonical' family of aerosols is seen here as essential to improve the understanding of their physical properties. The main goal here is the production of aerosols sufficiently stable for the realization of reference experiments to bring new elements in the understanding of (i) the dynamics of evaporation/condensation of droplet populations, (ii) the coalescence of droplets in turbulent flows, (iii) the influence of relative humidity in nucleation phenomena.

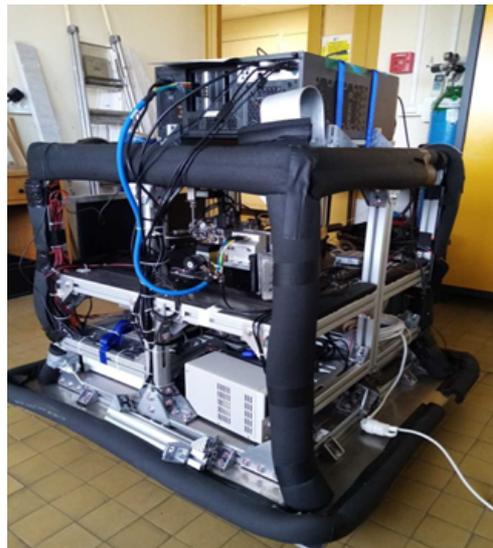

**Figure 1** : Overall view of the rack used for the parabolic flight experiments.

Three parabolic flight campaigns have been carried out, all aiming at validating the technical options implemented and at establishing the good feasibility of the experiments. So far, they demonstrated the possibility of producing populations of droplets by a protocol involving a fast compression of the air followed by a fast expansion (see sections 1 and 4) and validated aerosol observability by optical tomographic microscopy. The key point here was to ensure that the fast flows created by the compression/expansion sequence were damped over times smaller than the driving times of the physical phenomena governing the evolution of the aerosol when evolving in microgravity. The continuous phase being air, the appearance of long-lasting transient regimes was to be feared. Such behaviors would have been incompatible with an analysis by optical tomographic microscopy due to coherence loss between successive tomographic images. Preliminary estimates based on the numerical integration of a simplified model (a hard sphere with given initial velocity evolving in a continuous viscous phase) suggested damping times of the order of a few seconds. But this basic model was not accounting for important parameters such as the multiplicity of neighboring droplets, the finite geometry of the experimentation chamber (Figure 2), the pressure difference used for the expansion, etc. The parabolic flight experiments conducted in the frame of this work allowed to validate the dimensioning of the different elements of the equipment as well as the protocols to be implemented for the creation of the aerosols. They also evidenced typical damping times which turn out to be shorter than few seconds. Beyond these times, aerosols show up as droplet populations whose movements are only due to the residual acceleration of the flight conditions and not to the initial compression/expansion procedure. Sequences of about 15 seconds of reduced gravity were carried out during which the aerosols present a good stability thus allowing to track the evolution of several hundreds of droplets.

1. **Experimental cell and aerosol production protocol**

Aerosols are produced from moist air with a pneumatic circuit consisting of two chambers (Figure 2): a cylindrical experimentation chamber (EC, diameter 3 cm, height 6 cm) inside which droplets are generated and an expansion-compression chamber (ECC, volume of about 170 cm$^3$) acting alternatively as a gas reservoir or an expansion tank (Figure 2 and Figure 3). EC and ECC are initially at room temperature and pressure. They both contain air which may or may not be saturated with water. A membrane pump is used to circulate the air from the ECC into the EC. Then, by means of a system of valves, a fast expansion between the EC and the ECC is achieved. This expansion leads to a temperature drop inside the EC, which results in a decrease of the equilibrium saturation vapour pressure. An excess of water vapor is created that condensates into micron-sized droplets by nucleation[13][14][15].

Each experiment follows the same protocol: (i) closure of all valves, (ii) opening of TEV, (iii) starting up the pump for pressure increase in the EC up to a target value, (iv) closure of TEV, (v) few seconds delay for temperature relaxation, (vi) opening of PEV for air expansion and droplet formation, (vii) scanning tomography with image recording. The interest of this protocol is to maintain temperature and pressure in the EC before and after droplet formation almost constant and thus to avoid the implementation of an accurate thermal control of the whole set up which would be, from an instrumental point of view, highly challenging. This protocol takes only a few seconds and has the great advantage of limiting heat transfers especially with the walls of both EC and ECC.

A temperature measurement device composed of a type K thermocouple driven by an Agilent 34972A LXI is implemented inside the EC. The pressure inside the EC and the ECC are recorded with two pressure sensors (KELLER PAA-33X). A hygrometric sensor is mounted inside the EC (Sensirion SEK-SHT35). It also gives the temperature which allows a redundancy of the measurements. All the above sensors are placed at selected locations to provide measurements as accurate as possible without being too intrusive.

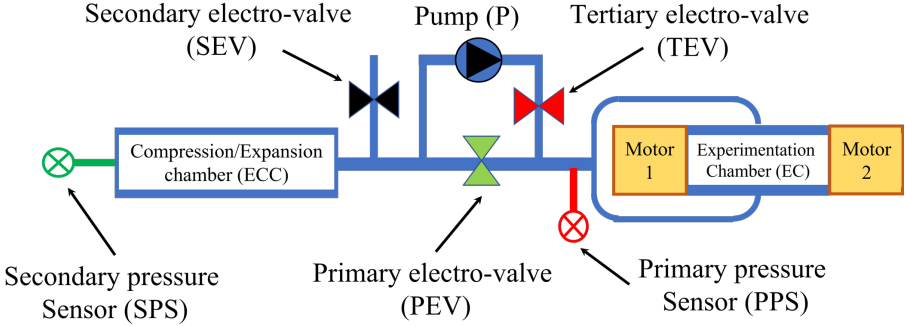

**Figure 2** : Schematic view of the experimental set-up.

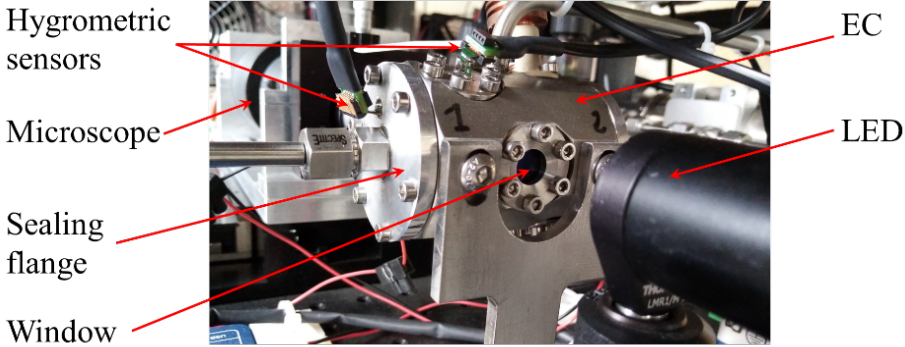

**Figure 3** : Photograph of the EC assembly. LED (foreground), EC (foreground), microscope (background). In this picture the motors used for Von Karman swirling flow generation are not mounted.

## 2. Optical tomography microscopy and imaging technique

Optical tomography microscopy in transmission mode is used to observe the droplets once formed and to track their evolution. This technique consists of acquiring images with a microscope with moving object plane. It has the advantage of enabling analyses in the visible spectrum and, in the particular case of optically transparent droplet populations as the ones investigated here, to allow in-depth analyses. The acquisition of tomographic scanning sequences allows a 3D reconstruction and thus a precise follow-up of the evolution of droplets over time. This technique is well known and has been for long used in medical fields[16][17][18] and, more recently, to describe emulsions[19][20]. An Optronis CP70-1HS-MC-1900 grabbing camera is used and mounted on a motorized translation stage (Figure 4). It allows a frequency acquisition up to 2000 fps in full frame that ensures reasonable coherence between all successive images. The microscope traveling distance is 2 mm with travel speed up to 10 mm/s. A tomographic shot (i.e. a single scan forwards or backwards) therefore lasts 0.2 s and generates 400 images of about 1 Mb each. The distance between two successive images is typically 5 µm. Tomography is applied perpendicularly, at ±1 mm on both sides of the symmetry axis of the EC and almost at its center. The overall analyzed area corresponds to a volume of about 2 mm$^3$.

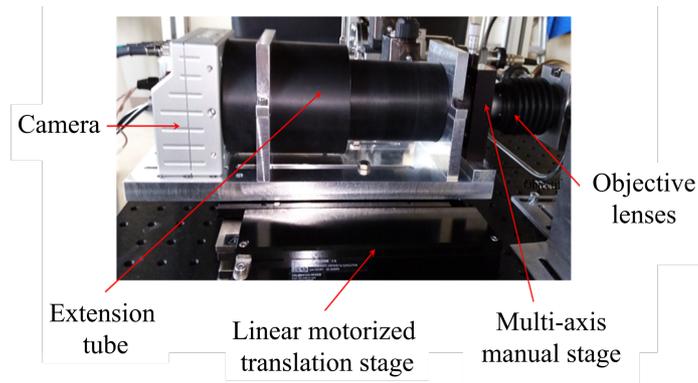

**Figure 4** : View of the microscope and optical tomography system.

Images are encoded on 8 bits and correspond to a field of view of 1280 µm by 860 µm (Figure 5). They all present a sufficient contrast (on at least 128 gray levels) for reliable post-processing. Vignetting may remain despite the optical quality and the collimation of the incident light beam. It is eliminated by subtracting a reference image acquired before the aerosol formation. The identification of the droplets and their location in the images is performed by thresholding predetermined gray levels (about ten). An image of the contours generated by each droplet is then obtained (Figure 6). The spherical geometry of the droplets and the fact that they are distant from each other produces typical circular contours, which is very helpful for the accuracy and CPU time effort of image post-processing. The principle of the imaging technique is to first detect the location of the center of each of the contours (Figure 6(b) and (d)), regardless of whether the corresponding droplet is focused or not, and then to construct, from the set of the centers of all the droplets and all the images of a given scanning shot, the individual optical axis of each droplet (Figure 7). The quality of the optics and image contrasts allows the detection of droplets with a radius down to 2 µm.

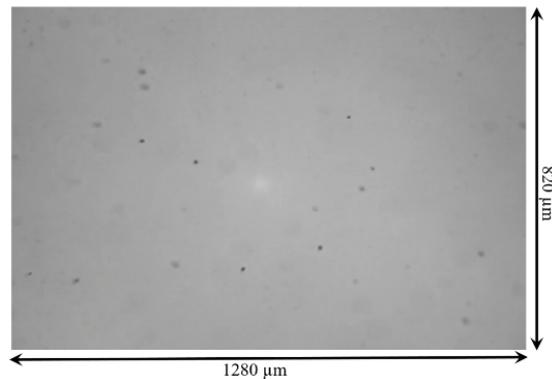

**Figure 5** : Typical image of a population of droplets. Droplet show up as black spots.

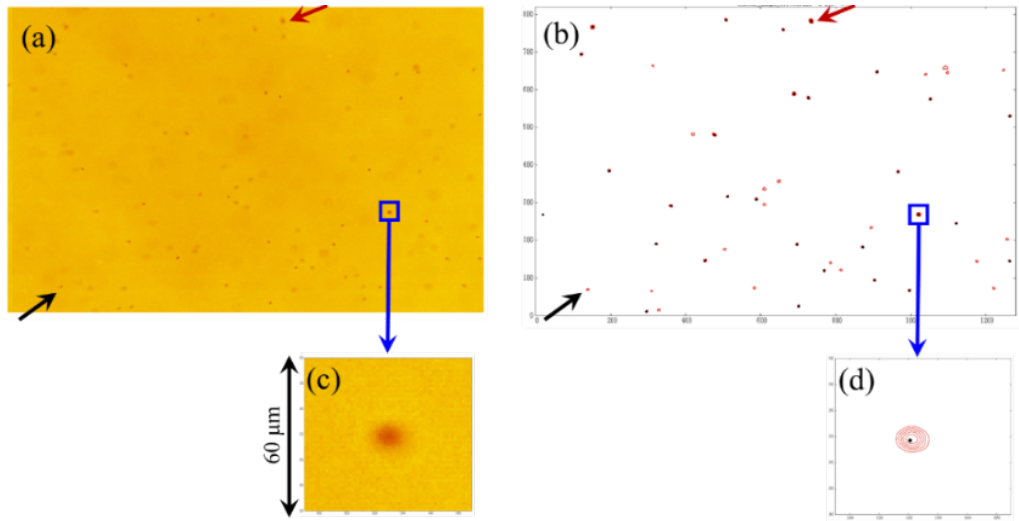

**Figure 6** : Image after background subtraction (a) and after treatment (b). The black and red arrows indicate two droplets and their signature after treatment. (c) and (d) correspond to a zoom on a droplet and both illustrate the gray levels produced by a droplet of about 10 µm in diameter. The black point is the center position obtained when averaging over all the contours.

Optical axis are processed one by one in a way to determine the droplet location in the scanning direction. This process is iterated for all the droplets (Figure 8). Optical axis construction requires coherence between consecutive images and thus an aerosol evolving on slow characteristic times when compared to those of the image acquisition system. This constraint is not satisfied right after droplet formation because of the fast flows generated by the initial expansion. As these flows decay after a few seconds (see below), sequences of several tens of thousands of coherent images are produced and therefore usable for post processing. Figure 7 represents all the optical axes for a single tomographic sequence. It shows hundreds of optical axes and illustrates the accessible precision of the analyses. This imaging technique is further used to obtain 3D representations (Figure 11).

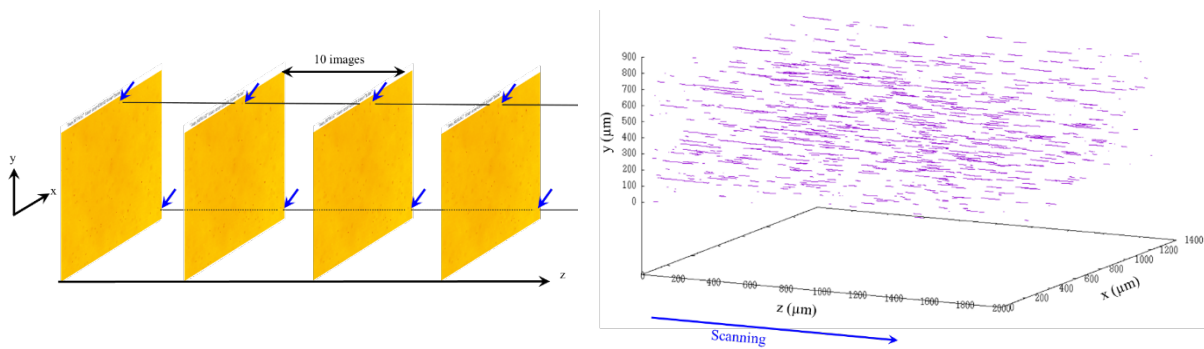

**Figure 7** : Left : Schematic illustration of droplet tracking. The two horizontal lines correspond to the optical trajectory of two droplets indicated by the blue arrows. The decimation of 10 images is used. The actual distance (resp. time interval) between each image of this figure is thus about 50 µm (resp. 5 ms). The coordinates (x,y) are in the plane of the images and z indicates the position of the microscope working plane. Right: Treated tomographic sequence. The orientation of the optical axes along the z the direction is a signature of the aerosol stability and good image coherence.

The identification of the optical response produced by each droplet leads to optical axes of length depending upon their diameter. The larger the radius of droplets, the longer their optical axis. Figure 8 illustrates this fact in a representation showing the radius obtained from the optical response of each droplet. When the working plane of the microscope is in the vicinity of a droplet location, the measured apparent radius reaches a minimum value. This property is used in the last phase of the image analysis procedure to estimate the radii of the droplets. The apparent radius of a given droplet is obtained from the value at half-height of the radial average of the gray levels of its optical response. This radius is larger when droplet is out of focus and minimal when it is in focus (Figure 8(b)). Its actual radius is finally computed from the minimum of a quadratic fitting of all the apparent radii when running along its optical axis. At this minimum contrast is the sharpest. Beside droplet radius, this procedure also determines the location of the droplet along its optical axis. This technique may however be too rough regarding to the comparatively large fluctuations shown Figure 8(b). Accurate measurements of droplet size distributions will therefore require further efforts and most likely the implementation of refined techniques such as FFT-based analysis.

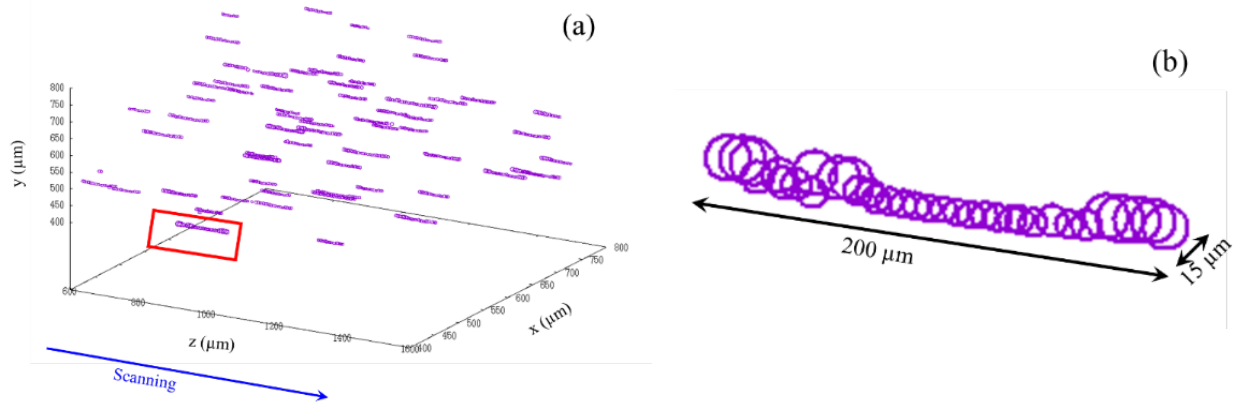

**Figure 8** : (a) Zoom into the domain 400<x<800 (µm), 400<y<800,600<z<1600 (µm) of Figure 7. The apparent radii of the droplets are shown as circles. (b) Evolution of the apparent radius of the droplet in the red frame of (a) as a function of the position of the working plane.

### 3. Set up for Von Karman swirling flows and droplet coalescence investigations

The EC is equipped with two coaxial counter-rotating rotors consisting of smooth flat disks of 29 mm diameter facing each other. They're used to generate a Von Karman swirling flow within the EC[21][22][23]. When activating the rotors, the air close to the disks is driven radially outward by the centrifugal force. This generates an axial flow toward the disks along their axis and a recirculation along the opposite direction along the EC lateral boundary. Therefore, a radially inward flow is generated in the midplane between the disks such that there is a stagnation point at the center of the EC. It is expected that the droplets, being subject there to the azimuthal velocity shear, are entrained by the air flow and accumulate in the vicinity of this stagnation point. The main interest of setting the droplets in motion in such a flow is to force them closer to each other. Such experiments are actually expected to help to understand coalescence of droplets within clouds[24] as they will give access to the evolution of the density (actually the liquid fraction) and the mean size of the droplets in the vicinity of the stagnation point. Investigations of the influence of the properties of the Von Karman swirling flows (VKSF), that can involve more or less turbulent regimes depending on the Reynolds number, are also planned.

The motion of the rotors is achieved with two FAULHABER servomotors (4221G 024EXTH) equipped with an encoder (IE3-2024) each controlled by a motion controller (MC5005 S RS) allowing regulation of the rotation frequency (motors are shown in Figure 2 but not in Figure 3). Rotors are both attached to a relay magnetically driven by a motion transmitter connected to the motors allowing rotation frequencies in the range 1 to 75 Hz. The sealing is obtained by a flange which separates the relay from the EC and thus from the outside. Magnetic neodymium segmental arc magnets provide the magnetic coupling. The main advantage of this configuration is that it allows for optimal sealing since no direct drive shaft has to be used.

### 4. Ground experiments

Ground experiments are essentially aimed at qualifying and understanding the response of experimental devices and to ensure control of experimental conditions in terms of pressure, temperature and relative humidity. For all experiments presented in this section, initial pressure and temperature are ambient ones while relative humidity (H) ranges between 2% and 96%. The aerosol formation protocol (section 1) and the optical tomography analysis are synchronized, which allows for a fast triggering of all devices (actually with a single 'click'). This is necessary because of the fast gravity drainage of the droplets and, for parabolic flights, the limited duration of the microgravity periods.

Figure 9 displays the temperature and pressure evolution when producing an aerosol with a target pressure of 2000 mbar. Reference time (t=0) is set at the opening of PEV (step (vi)). As expected, two jumps in both temperature and pressure obviously associated to step (iii) and (vi) show up. Temperature relaxes back to its initial value sufficiently fast for both steps which confirms that air, within the EC, keeps close to its initial value despite droplet formation. In other words, forming droplets is not expected to yield significant heat transfers with the boundary conditions (in particular the walls of EC). It is therefore reasonable to assume that aerosol production can be achieved, at first approximation, under quasi-isothermal conditions.

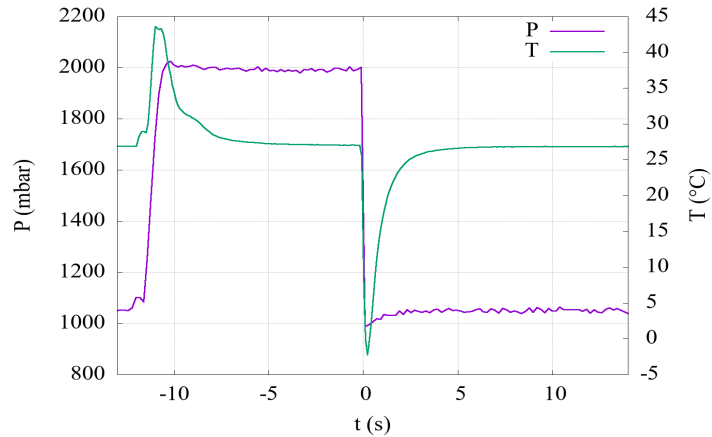

**Figure 9**: Evolution of pressure and temperature as a function of time inside the EC. Reference time t=0 is set at the opening of the PEV (step(vi)).

The typical temperature relaxation time, $t_{relax}$, as a function of relative humidity H is plotted in Figure 10 for 108 experiments and four target pressures (2000, 2100, 2200 and 2300 mbar). $t_{relax}$ is defined as the time required for the system to recover 95% of its initial temperature after step (vi). This figure shows that the higher the relative humidity, the higher the relaxation time. This is a straightforward consequence of the larger thermal heat capacity of water loaded air. The increasing trend of $t_{relax}$ as a function of H is therefore consistent with thermodynamics and a straightforward consequence of the fact that equilibration time of a system increases with its thermal capacity. The set of experiments of Figure 10 also provides fundamental information about droplet formation. Image analysis (not shown here) indicates indeed that within the pressure ranges investigated here, no droplets are formed when H < 40%. It also appears that the overall number of droplets (resp. the typical radius of the droplets) increases (resp. reduces) when increasing the target pressure. This turns also consistent with thermodynamics but, droplets getting drained very fast due to buoyancy forces, reliable quantitative investigations are not possible.

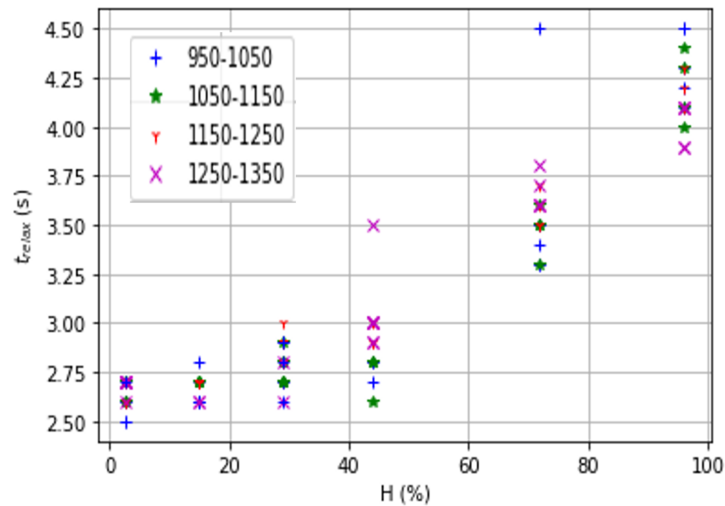

**Figure 10** : Time required for temperature stabilization after expansion as a function of H. The legend indicates the target pressure range in mbar at step (iii) (compression phase). For these experiments, temperature is 24°C ± 2°C.

## 5. Microgravity preliminary results

All the results presented in this section were obtained from three CNES parabolic flight campaigns carried out in October 2020, March 2021 and March 2022 in Bordeaux on board the Airbus A310 0G managed by Novespace. Aerosol formation follows the same protocol as for ground experiments. The pressure in the EC is well controlled and fixed at 1 atm. It is thus slightly higher than that of the cabin. The conditions in temperature are determined by the temperature in the cabin and are thus not controlled. It tends to increase as the experiments are carried out but remains within the range of 18 and 28°C, over the duration of each flight. This temperature drift obviously results from the activity of the people and the numerous experiments running onboard. Given the precision of the measurements and the way experiments are planned for each parabola sequence, this problem is not expected to significantly affect the accuracy of the experimental results.

The first fundamental question that experiments had to answer was the possibility of producing aerosols and to verify that they were stable enough to maintain coherence between the tomographic images for relevant image post-processing. All experiments have clearly demonstrated the possibility of producing droplets. The continuous phase being air, fast turbulent transient regimes occur right after expansion (step (vi)). This makes tomographic microscopy analysis potentially ineffective, as image coherence is essential. Fortunately, it turns out that the rapid flows created by the initial expansion are damped out within typically 4 to 7 seconds. Beyond this time, droplets exhibit small amplitude movements only due to the residual acceleration of the parabolic flight conditions. Periods of about 15 seconds of reduced gravity are thus accessible during which their evolution can be studied under scientifically relevant conditions. Figure 11 shows an example of a 3D reconstruction of a single tomographic shot and illustrates the achieved level of precision.

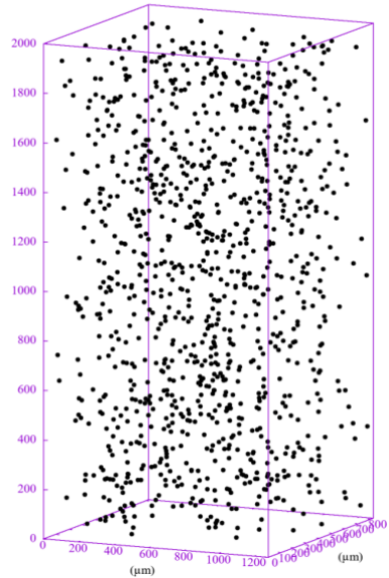

**Figure 11** : 3D reconstruction of an aerosol. Droplets are black spots. Their radius is multiplied by a factor 10 for visibility. About 24 hours CPU are necessary for such a reconstruction on a basic work-station. This experiment was performed in the CNES 59$^{th}$ parabolic flight campaign (parabola 57, September 2020) with and EC diameter is 4 cm and height 4 cm.

Tomographic shots are processed one by one to identify each individual droplet. This process is iterated to investigate the properties of the droplet populations generated for each experiment. Figure 12 shows the evolution their number (N) as a function of time for four replica experiments and two values of relative humidity: water saturated (H=100%) and sub-saturated (H=70%). The expansion phase starts at t = 0 and is operated at the very beginning of the parabola microgravity phase. The gray shaded area indicates the time period within which image reconstruction quality is poor due to the lack of coherence between successive images. Here, it lasts for typically 7 s. Beyond this time, water-saturated experiments show droplets persisting for the remaining duration of the parabolas while they have all disappeared after typically 11 s when H=70% [25][26].

These results clearly demonstrate the possibility to measure the number of formed droplets in microgravity conditions. This is an essential condition for assessing the feasibility of aerosol investigations over longer periods of time. They further indicate that evaporation of droplets in the sub-saturated regime is a very fast phenomenon as it takes place in less than 0.1 s. Such investigations could for sure complete the very rich literature focused on the evaporation of sessile droplets in microgravity[27][28][29]. Furthermore, fundamental questions such as how quickly droplet populations evaporate as relative humidity conditions change or the precise number of droplets (e.g. liquid fraction) remaining after expansion become quantitatively accessible. But Figure 12 raises practical questions that have yet to be resolved. The experiments show indeed poor reproducibility. Broader ensemble of data have to be collected to reduce the statistical variability of probing a small volume. Another explanation could be associated to unexpected differences, up to couple of seconds, in the duration of step (v) of aerosol formation. This problem results from the lack of precise synchronization in the operation of the pump (for pressure increase in the CE) and the valves opening/closing (for aerosol formation). The consequences of this uncontrolled delay are most likely differences in the pressure drop during aerosol formation and thus modified heat exchanges from one experiment to another. This instrumental problem has been solved and forthcoming parabolic flight experiments are expected to show a significantly improved reproducibility.

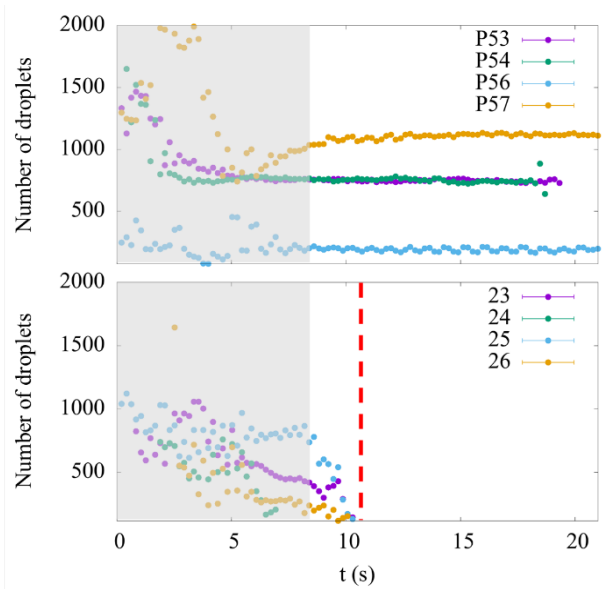

**Figure 12** : Evolution of the number of droplets as a function of time for target pressure 2000 mbar and temperature 22 °C. Top: H=100%. Bottom H=70%. The gray shaded area corresponds to the transient evolution phase. Vertical red dashed line indicates evaporation time. For parabolas P53 and P54 microgravity duration was shorter than expected. These experiments were performed in the CNES 59[th] parabolic flight campaign (September 2020). For both figures, EC diameter is 4 cm and height 4 cm.

Parabolic flights have demonstrated that it is possible to generate stable aerosols only subject to the residual acceleration of flight conditions (Figure 11). Another fundamental point to check here is the ability of the rotor device presented in section 3 to generate a VKSF within the EC and to set the overall aerosol in motion. The rotors have a smooth surface (not equipped with blades) and the question is whether it is possible or not to make droplets move and, if set in motion, to determine the time required for the establishment of a VKSF. Tuning parameter is the rotation frequency (noted f hereafter) imposed on the rotors. In practice, it is not possible to have a global view of this flow as only a very limited volume in the vicinity of the center of the EC is analyzed (section 2). But, in view of its structure, this volume is expected to contain the stagnation point in the vicinity of which the droplets are expected to gather. This phenomenon is unfortunately not observable in the experiments because the periods of microgravity are too short and the positioning of the rotors most likely lack of precision. Still, the droplets, once formed, can be used as tracers. The analyzed area being in the center of the EC, it is reasonable to consider that the VKSF is established in the whole EC once they're set in motion (actually transported by the VKSF).

The experiments are focused on the measurement of the drag time ($t_{dr}$) corresponding to the time lag between rotor activation and droplet transport by the VKSF. The protocol is similar to that of section 1 with the following additional step: after 7 seconds of an ongoing parabola, as the droplets are at rest, the rotors are activated. The drag time $t_{dr}$ is evaluated by a direct observation of the movement of the droplets. Its value is determined when they appear to exhibit a coherent motion over a distance larger than 100 µm. Such movements are easy to identify by eye and cannot be the consequence of residual acceleration. Results are plotted in Figure 13 and indicate that $1/(f \times t_{dr})$ increases linearly with f in the investigated frequency range ($6 \leq f \leq 15$ Hz). It extrapolates down to zero for $f \approx 1.5$ Hz which is consistent with the fact that no motion is observed for the duration of the parabolas when $f < 4$ Hz. This corresponds to a Reynolds number of the order 400 ($Re = 2\pi f R^2/\nu$, where $\nu \approx 1,5 \cdot 10^{-5}$ m$^2$.s$^{-1}$, is the kinematic viscosity of air and R ≈ 1.45 cm de rotor radius) for which it is known that the first instabilities of the VKSF take place. For smaller values the flow is laminar and the motion develops on a diffusive time scale $(2R)^2/\nu \sim 60$ s after the rotors are set into motion. This turns longer than the duration of the parabolas. No flow is therefore observable.

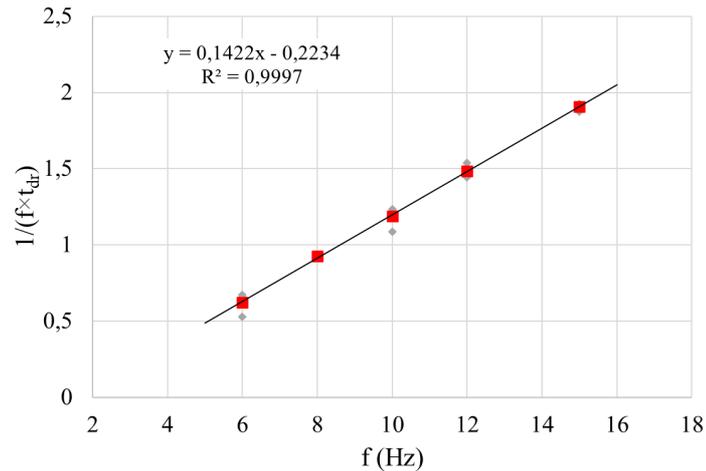

**Figure 13** : $1/(f \times t_{dr})$ as a function of rotation frequency f. Five replica experiments are achieved for each experiment (small gray diamonds). The red squares are the average value for each replica series.

**Conclusions**

The developments carried out within this work for the study of the populations of water droplets in microgravity required the implementation of specific instrumental techniques. Parabolic flight experiments were performed to verify their relevance. For the formation of the droplets, a protocol consisting of compressing and then expanding the air in a cylindrical experimentation cell was implemented. This protocol is short enough to assume droplet production to be a quasi-isothermal process. Preliminary parabolic flight experiments have established the possibility to quantitatively describe droplet populations under microgravity conditions. The flows generated by the initial expansion fade away in few seconds. Beyond this time, as residual acceleration is small, the droplets are almost at rest. Optical microscopy tomography in transmission mode is used to reconstruct the droplet populations in 3D and to follow their evolution in time over more than 15 seconds of microgravity. As the thermodynamic conditions are controlled, the evaporation dynamics of large droplets can be investigated. This work also shows how droplets can be set in motion by creating a Von Karman flow. The instrumental developments will be continued in the next future. Important improvements remain to be undertaken for the control of the relative humidity and the thermal regulation of all the elements containing the air used for the droplet formation.

**Acknowledgements**

Authors gratefully thank CNES, in particular Christophe Delaroche and Laurent Arnaud, CNRS and Région Provence-Alpes-Côte d'Azur for important financial support, ESA - European Space Agency, within the project ESA/EVAPORATION (ESA Contract Number 4000129506/20/NL/PG) and Yvan CECERE from MADIREL for technical support.